\documentstyle[11pt,epsf,psfig]{article}

\begin{document}
\newcommand{\wavn}{\mbox{cm$\rm ^{-1}$}}
\newcommand{\nuone}{\mbox{$\nu_1$}}
\newcommand{\nutwo}{\mbox{$\nu_2$}}

\newcommand{\HH}{\mbox{H$\rm _2$}}
\newcommand{\HHH}{\mbox{H$\rm _3$}}
\newcommand{\HHHp}{\mbox{H$\rm _3^+$}}

\newcommand{\DD}{\mbox{D$\rm _2$}}
\newcommand{\DDp}{\mbox{D$\rm _2^+$}}
\newcommand{\DDD}{\mbox{D$\rm _3$}}
\newcommand{\DDDp}{\mbox{D$\rm _3^+$}}

\newcommand{\twopE}{\mbox{2p$\rm \ ^2E'$}}
\newcommand{\twosA}{\mbox{2s$\rm \ ^2A_1'$}}

\newcommand{\twop}{\mbox{2p$\rm \ ^2A_2''$}}
\newcommand{\twopNK}{\mbox{2p$\rm \ ^2A_2''$(N=K=0)}}

\newcommand{\thrsA}{\mbox{3s$\rm \ ^2A_1'$}}
\newcommand{\thrsANK}{\mbox{3s$\rm \ ^2A_1'$(N=1,K=0)}}

\newcommand{\thrpA}{\mbox{3p$\rm \ ^2A_2''$}}

\newcommand{\thrdE}{\mbox{3d$\rm \ ^2E''$}}
\newcommand{\thrdENGR}{\mbox{3d$\rm \ ^2E''$(N=1,G=0,R=1)}}

\newcommand{\nsone}{\mbox{ns$\rm \;^2A_1'$}}
\newcommand{\ndone}{\mbox{nd$\rm \;^2A_1'$}}
\newcommand{\ndthr}{\mbox{nd$\rm \;^2E''$}}

\newcommand{\Aonep}{\mbox{A$\rm _1'$}}
\newcommand{\Ep}{\mbox{E$'$}}

\title{Vibrational Frequencies of the \twop\  and \thrdE\ States of the 
  Triatomic Deuterium Molecule }
\author{ U. M\"uller, M. Braun, R. Reichle, and
  R. F. Salzgeber \\
 Universit\"at Freiburg, Fakult\"at f\"ur Physik, \\
  Hermann-Herder-Str. 3,  D-79104 Freiburg, Germany}
\date{\today}
\maketitle

\normalsize
\begin{center} Abstract \end{center}

We investigated the vibrational energies in the \twop\ and \thrdE\ 
states of  the triatomic deuterium molecule \DDD .  
The experiments were performed
using a fast neutral beam photoionization spectrometer recently developed at 
Freiburg. 
A depletion type optical double-resonance scheme using two pulsed dye lasers
was applied. 
The measured vibrational frequencies of the \twop\ state of \DDD\ are compared 
to those of \HHH\ and to 
theoretical values calculated from an {\it ab initio} potential energy surface. 
The data give insight into the importance of the coupling between the valence
electron  and the ion core.

\vspace*{1cm}
PACS-Numbers: 33.80.Rv, 33.20.Kf, 33.20.Lg, 34.10.+x, 34.30.+h 

\newpage
\section{Introduction}

The triatomic hydrogen molecule, the simplest neutral polyatomic molecule, is
ideally suited to study fundamental aspects of the interaction between the
electronic and the nuclear motion. 
The molecule is sufficiently small for high precision
{\it ab initio} investigations of the potential energy surfaces to be
successful \cite {NJ82,PKKW95}. 
Excited states of
\HHH\ and
\DDD\ were discovered by  Herzberg and coworkers
\cite{HerzbI,HerzbII,HerzbIII,HerzbIV}  in emission from  electrical gas
discharges in hydrogen and deuterium.  All of the excited states are
predissociated  by coupling to the repulsive 
ground state potential energy surface except for the single rotational state
\twopNK\   which 
is metastable  and can be prepared in a fast neutral beam by charge
transfer of \HHHp\ or \DDDp\
\cite{HLCH89,Dev69,GP83,GK84} in cesium.  The  \HHH\ \twopNK\ state was 
used as a platform for laser-excitation/ionization experiments  in order to
explore the  Rydberg states of \HHH\ and to determine their vibrational
frequencies \cite {HLCH89,He86,He88,DKMW88,KMW89,LPH88,LH89,LHH89} .

In a recent investigation from this laboratory \cite{MMRB97}, we created a 
fast  beam of rotationless \DDD\ molecules in the lowest
vibrational levels of the
\twop\ metastable state.  Following vibrationally diagonal 
excitation in the ultraviolet spectral range, we observed five Rydberg series. 
Two series were detected by field-ionization 
and assigned to  s-- and d-- type Rydberg states 
with a vibrationless \DDDp\ core.  Three series were found to consist of 
 transitions from vibrationally symmetric stretch and degenerate mode excited
\twop\ states into  auto-ionizing Rydberg states. In order to assign the 
core vibrational excitation, we used preliminary results of 
a depletion type double resonance  experiment.
A vibrationally non-diagonal transition which
appears as a Fano type line above the first ionization limit allowed us to 
 determine the first symmetric stretch 
vibrational frequency of the \DDD\ \twop\ state. In order to determine 
the degenerate mode frequency of the \twop\ state, we used  the molecular 
constants of the \DDDp\ ion\cite{ACCK94}.

The objectives of the present investigation are threefold. First, we present 
the results of double resonance experiments which confirm  the 
 assignment of the Rydberg states observed previously \cite{MMRB97}. 
Secondly, we
investigate vibrationally  diagonal and non-diagonal transitions from the
\twop\ state to vibrationally ground and exited levels of the \thrdE\ state. 
In this way,  we determine the vibrational
frequencies of the \twop\ and \thrdE\ states of \DDD . 
Third, we calculate the
vibrational levels of the \twop\ states of \HHH\ and
\DDD\ with the potential energy surface published recently by Peng et al.
\cite{PKKW95} using the filter diagonalization method 
in the version of Mandelshtam and Taylor
\cite{mandelshtam:97a}. The comparison between the experimental data and
the theoretical model gives insight into the influence of the 
Rydberg electron on
the bond strength. By using different isotopes, the potential energy
surface   is effectively probed.

\section{Experimental}

In this study, we used the fast beam collinear spectrometer 
recently developed at Freiburg \cite{MMRB97}. 
Triatomic deuterium ions  (\DDDp ) were created in a 
hollow cathode discharge in  
deuterium  (\DD ). The cathode was  cooled 
by liquid nitrogen. The ions were accelerated to an energy of 3.6 keV and
mass selected by a Wien-filter. 
A small fraction of the ions  was neutralized by
charge transfer in cesium vapor. After the charge-transfer cell, 
the unreacted ions were removed by an electric field. 
A 1 mm diameter aperture located 30 cm downstream of the 
charge transfer cell was used to stop products of dissociative charge
transfer. Fast metastable molecules which entered 
the 120 cm long laser-interaction region were excited 
by a counterpropagating  pulsed dye laser beam. 
Photoions were detected at the end of the interaction region 
by an energy analyzer and a microsphere plate.
A 200 MHz dual-counter was used to accumulate the signal produced by the laser
pulses  and the background events  separately.   The data were then 
transferred to a laboratory computer and stored for further treatment.

In order to perform depletion type double resonance experiments, 
we operated two dye lasers 
 pumped by an excimer laser. 
The wavelengths of both dye lasers  were programmed and/or scanned 
under control of the laboratory computer. 
Using an optical delay line, the pulses of one of the  dye lasers
(labelling laser) were delayed by about 15 ns with respect to the pulses of 
the other dye laser (excitation laser).  Both laser 
beams were merged by a beamsplitter before they entered the interaction
region.  
The labelling laser  was operated at a
fixed wavelength, and used to excite a transition from a specific
vibrational level of  the \twop\  state  to a field- or autoionizing
state. The ion signal was used to monitor the population in the  lower
state. The excitation  laser was scanned in the tuning ranges of the  dyes
Rhodamine 6G and Coumarin 307. 
Transitions originating from 
the labelled lower state lead to a reduction of the ion signal 
 and  appear as  depletion dips in the spectra. 
The depletion technique is most effective in the case of transitions to 
upper states which are quickly predissociated. A long-lived
upper state may be ionized by a second photon from the 
same excitation laser pulse (1+1 REMPI) which reduces the depth 
of the depletion feature. In that case, the intensity of the excitation laser 
was reduced by neutral density filters. 

In order to calibrate the wavelength scales of the lasers,  the
optogalvanic signal from a hollow cathode discharge in neon and argon  was
recorded, and the observed lines were compared to tabulated values
\cite{MIT-tab}.
The Rank-formula \cite{Rank59} was 
used to correct for the refractive index of air. 
The Doppler-shift due to the motion of the fast neutral 
molecules was calculated from the acceleration voltage of the ion source. 
We estimate the systematic uncertainty of the photon energy scale 
to be less than  0.2 \wavn .

\section{Numerics}
The significance of any molecular eigenstate calculation, depends strongly
on the quality of the potential energy surface (PES) on which it is performed.
For computing the low lying vibrational frequencies of H$_3$ and D$_3$ in the 
2p $^2$A$_2^{''}$ state, we use the fitted {\it ab initio} potential energy 
surface of Peng et al.\cite{PKKW95}. This seems to be a rather accurate fit
(with only 5 cm$^{-1}$ average deviation from its 1340 {\it ab initio} points 
\cite{PKKW95}), thus enabeling us to make a rather precise prediction of
its eigenstates.\\
>From the various methods for computing vibrational states of small molecular
systems, we use the low-storage version of the filter-diagonalization method 
introduced recently by Mandelshtam and Taylor \cite{mandelshtam:97a}. This 
method is conceptually based on the filter-diagonalization procedure of Wall 
and Neuhauser \cite{neuhauser:95} which extracts the system eigenenergies by 
harmonic inversion of a time correlation function C(t). The method of 
ref.~\cite{mandelshtam:97a} is designed to use a direct harmonic inversion of 
the Chebyshev correlation function \cite{tal-ezer:84}
\begin{equation}
c_n =\langle\xi_0|T_n(\hat{H})|\xi_0\rangle\sim\sum_k d_k \cos n\omega_k\ ,
\end{equation}
for the eigenenergies $E_k=\cos\omega_k$ and amplitudes $d_k$. The
computation 
of the $c_n$ sequence is done to essentially the machine precision using a 
very inexpensive iterative numerical scheme,
\begin{equation}\label{eq:cheb}
\xi_1=\hat H\xi_0,\ \dots\ ,\ \xi_{n+1}=2\hat H\xi_n-\xi_{n-1}\ ,
\end{equation}
with $c_n$ being generated using
$c_{2n}=2\langle\xi_n|\xi_n\rangle-c_0,\ \  c_{2n-1}=2\langle\xi_{n-1}|\xi_n
\rangle-c_1$.
This requires to store only a few vectors at a time, if the 
matrix-vector multiplication is implemented without explicit storage of the 
Hamiltonian matrix. The spectral analysis part (i.e., the harmonic inversion 
of $c_n$ by the filter-diagonalization) is carried out independently and 
efficiently after the sequence $c_n$ is computed. All these features imply
the 
very high performance of the overall numerical procedure.\\
It is not hard to achieve a precision of the low lying states of 
$\Delta E / E= 10^{-4}$, because their density of states is not 
high, and, therefore, the number of iterations can be small ($\sim 10^3$), 
requiring only a few hours of CPU time on a RS6000/59H workstation.\\
We choose Radau coordinates, in which all mixed derivatives in the kinetic
energy vanish \cite{sutcliffe:91}, to guarantee for a fast application of the 
Hamiltonian  to a vector, which is the bottleneck in iterative methods. 
However, this choice of coordinates implies that C$_{2v}$ symmetry, rather
than the full D$_{3h}$ symmetry of H$_3$ (D$_3$), is used in the
calculation. Nevertheless, E states can be identified as a numerically close
pair of an A$^\prime$ and an A$^{\prime\prime}$ state \cite{tennyson:85}.\\
We also use a sinc-DVR \cite{colbert:92} for the radial and a Legendre-DVR
for 
the angular part of the Hamiltonian. The parameters defining the grid are the 
size of the primitive sinc-basis $n_1 = n_2$, which are truncated by a
kinetic 
energy cutoff to $n_{1b} = n_{2b}$, their spatial extension from $r_{i\,min}$ 
to $r_{i\,max} (i=1,2)$, the size of the primitive basis of the Legendre 
polynomials for the angular motion $n_3$ and a three-dimensional potential 
energy cutoff $V_{cut}$. All primitive basis parameters have been adjusted in 
the one-dimensional problem, with the remaining coordinates held fixed to 
their equilibrium values. The details of the convergence procedure will be
described in \cite{salzgeber:97a}.\\ 
For $H_3$ we used the values $n_1 = n_2 = 71, n_{1b} = n_{2b} = 60, n_3 = 67,
V_{cut} = 13.06 eV$ above the minimum of the potential, and $r_{1\,min} = 
r_{2\,min} = 0.7 a_0, r_{1\,max} = r_{2\,max} = 7.1 a_0$, overall resulting
in 
232 705 gridpoints. The figures for $D_3$ were $n_1 = n_2 = 49, n_{1b} = 
n_{2b} = 43, n_3 = 67, V_{cut} = 12.06 eV$ above the minimum of the potential,
and $r_{1\,min} = r_{2\,min} = 0.8 a_0, r_{1\,max} = r_{2\,max} = 4.2 a_0$, 
resulting in 116 192 gridpoints.\\
The results, which are shown in table \ref{tb:VibSep}, 
have an accuracy of better than 0.25
cm$^{-1}$ assuming the PES is correct, 
which e.g. follows from comparison of the A$^\prime$ and  
A$^{\prime\prime}$ components of the E states. Additionally, 
convergence was ensured
by running various grids of increasing size. 

\newpage
\section{Results and Discussion}

In Table \ref{tb:Series}, the ionization limits  and 
quantum defects of  five
Rydberg series  detected in a previous investigation from this laboratory
\cite{MMRB97}  are listed. Series 1 and 2 were observed by field-ionization 
and  assigned to d- and s-type Rydberg series, respectively, converging
to a vibrationless \DDDp ($N^+=1,K^+=0$)  state.   
Series 3 to 5 were found to arise from autoionizing states which 
are built on vibrationally excited \DDDp\ ion cores.  
We also observed a Fano type resonance at 29652.6 \wavn\ 
in the continuum above the
first ionization limit of the vibrationless \twop\ state. 
 In the present investigation, the depletion technique is used in order  to
decide which transitions share a  common lower state, and to  determine
vibrational frequencies in a straightforward way. 

\subsection{Depletion measurements of the vibrationless \twop\ state}

\begin{figure} [h]
\psfig{figure=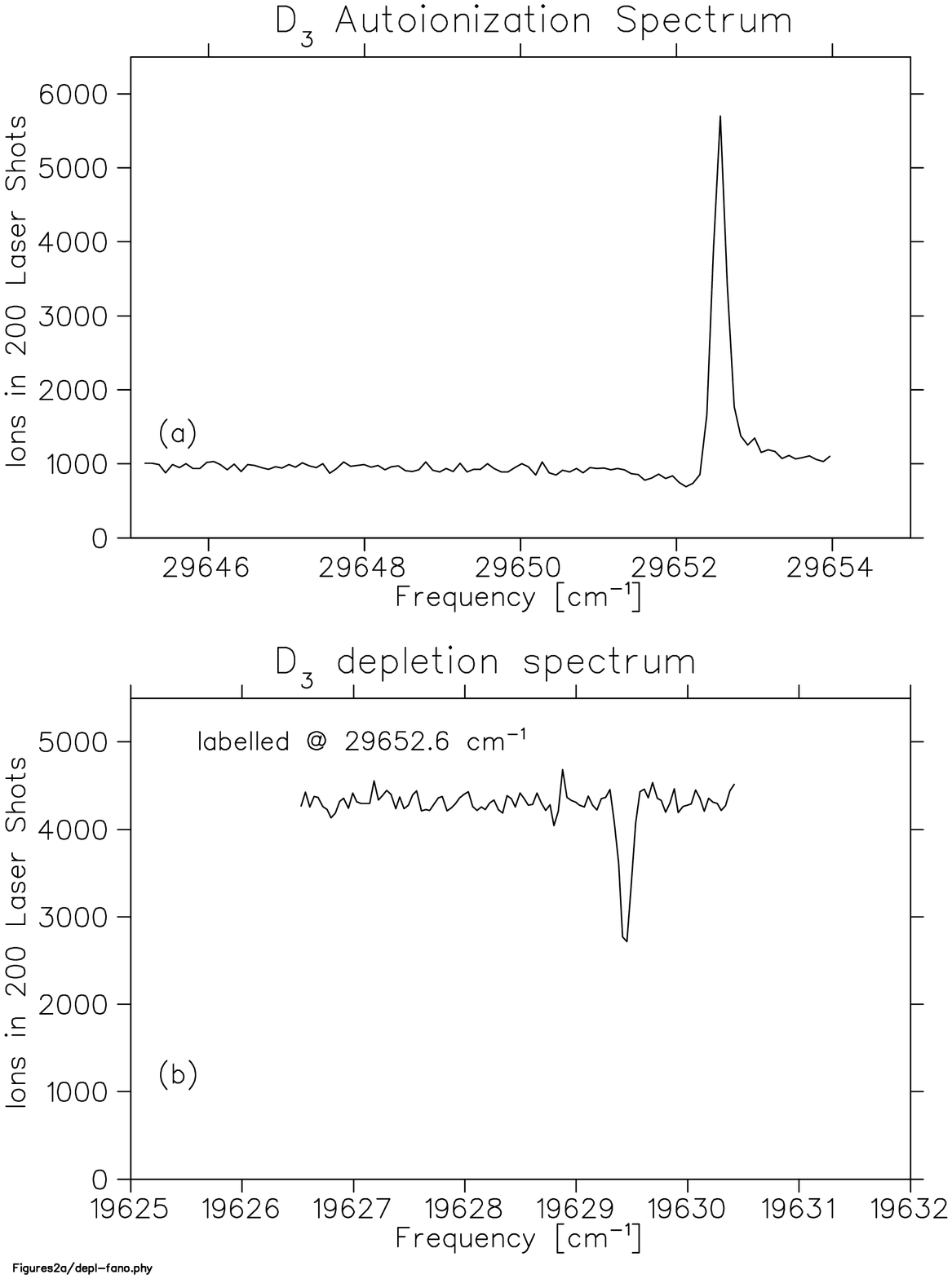,width=16cm}
\caption [Depletion Series1]
 {\label{fig:depl_3dn0n1} 
   Depletion spectra of  
   metastable \DDD\ molecules.
   The labelling laser was set to excite the n=39 (a) and n=40 (b)
   lines of the field-ionizing Rydberg series 1, respectively. 
  The excitation laser was scanned between 17329 \wavn\ and 17339.5 \wavn\ 
  (part a) in the tuning range of Rhodamine 6G, and 
  between 19623 \wavn\ and 19635 \wavn\ in the tuning range of Coumarin 307 
  (part b). 
 }
\end{figure}

 With the labelling laser 
set to a member (n=39) of the field-ionizing  series 1,  we observed a depletion
dip centered at a  frequency of  17333.3 \wavn (Fig. \ref{fig:depl_3dn0n1}a). 
This transition was observed in emission by Herzberg's group 
and assigned to the \twop (0,0,0,0) $\leftarrow$ \thrdE (0,0,1,0) transition of
\DDD \cite{HerzbIV}. We describe the rovibrational states by
the set  (\nuone ,\nutwo ,N, G) with  the quantum numbers of
 the symmetric stretch (\Aonep ) and the degenerate mode (\Ep ) 
vibrations  \nuone\ and \nutwo , the total angular momentum apart from 
spin $N$, and Hougen's convenient quantum number $G=l+\lambda-K$
\cite{Hougen62},  which contains the  projections of the total ($K$), the
electronic ($\lambda$), and the vibrational angular momentum ($l$) onto the
figure axis. The depletion feature in Fig.
\ref{fig:depl_3dn0n1}a shows clearly  that  the vibrationless \twop (0,0,0,0)
state is the lower state of Rydberg series 1 and 2. The large width of 
0.8 \wavn\  is mainly due to  power-broadening .  The depletion in the
center of the dip is almost complete which  indicates a strong decay mechanism
of the upper state either by predissociation or by radiation into a
state different from  \twop (0,0,0,0). 

We scanned the excitation laser in the tuning
range of Coumarin 307, and observed in a 300 \wavn\ wide region 
a single depletion
dip centered at 19629.4 
\wavn\ (labelling by n=40 of series 1, Fig.\ref{fig:depl_3dn0n1}b).
The width of the dip is about 0.2 \wavn\ FWHM.  
The  difference  between the locations of the depletion features shown in
Fig.\ref{fig:depl_3dn0n1}  is 2296.1
\wavn . This value is extremely close to the 2300.843 \wavn\ symmetric stretch
vibrational frequency of the \DDDp\ ion \cite{ACCK94}. 
We, therefore, assign the upper state of the transition at 19629.4 \wavn\ 
to the vibrationally symmetric stretch excited \thrdE (1,0,1,0) state. 
The Franck-Condon factor of this vibrationally non-diagonal transition is 
quite small.  
Therefore, the transition to the \thrdE (1,0,1,0) state (Fig.
\ref{fig:depl_3dn0n1}b) is much less
power-broadened than that to the \thrdE (0,0,1,0) state (Fig.
\ref{fig:depl_3dn0n1}a).  
The depletion depth  in Fig. \ref{fig:depl_3dn0n1}b is 
about half of the ion signal. With the exitation laser (Coumarin 307)
unattenuated and the ionization laser blocked, we  observe a strong REMPI peak
at 19629.4 \wavn . 
Both observations are in line with a  comparatively long 
natural lifetime  of the vibrationally excited \thrdE\ state. 

\begin{figure} [h]
\psfig{figure=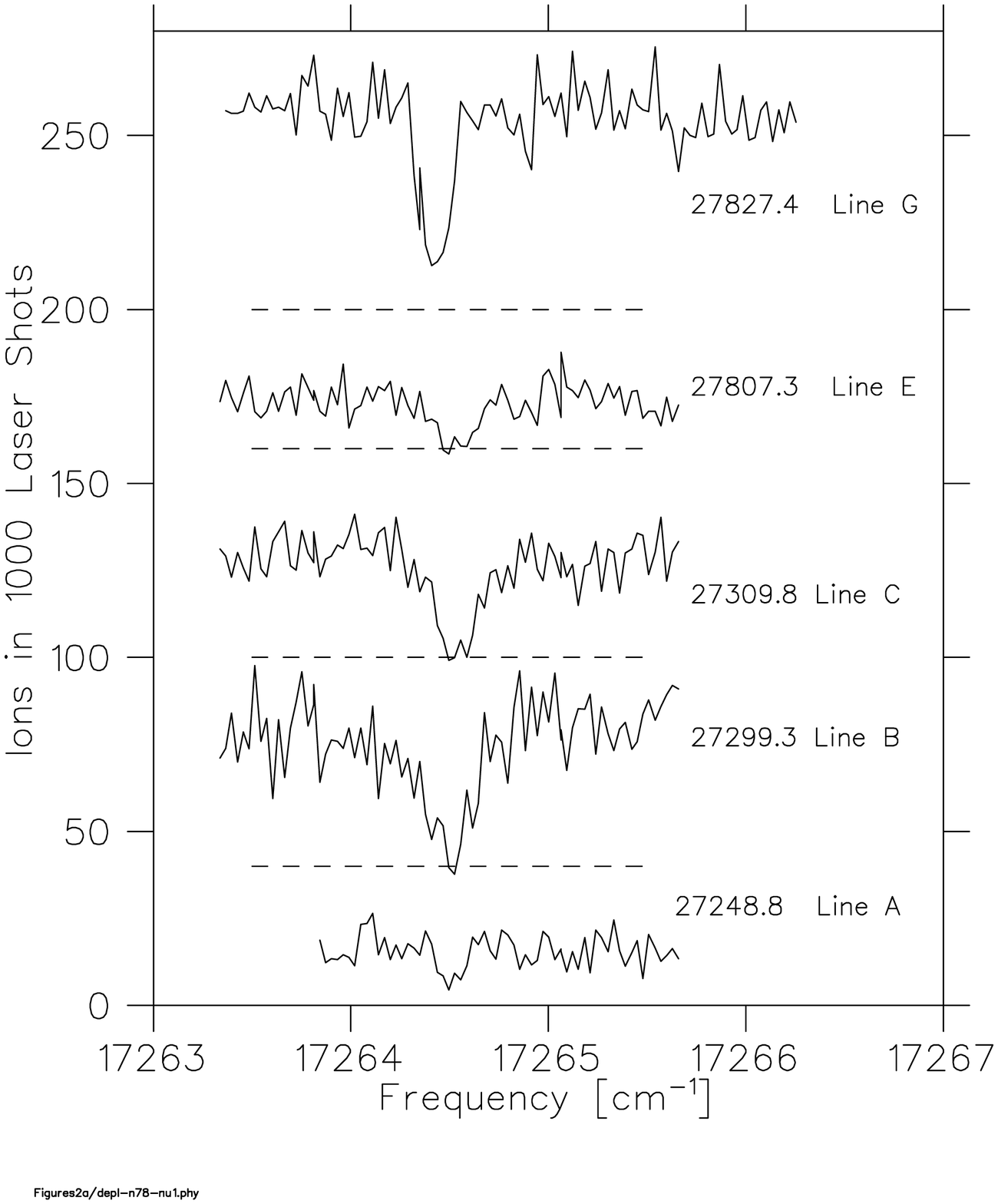,width=16cm}
\caption [Depletion Fano Peak]
 {\label{fig:depl_fano} 
   \\ (a) \ 
   Fano type resonance above the first ionisation threshold of the 
   \twop (0,0,0,0) state of \DDD . \\
   (b)
   Depletion spectrum  with the labelling laser set to the maximum of the 
   Fano  resonance at 29652.6 \wavn\ . The excitation laser was 
   operated with the dye  Coumarin 307 and scanned in the 19626.5 \wavn\ to 
   19630.5 \wavn\ frequency range. 
 }
\end{figure}

With the labelling laser tuned to the peak of the Fano type resonance 
at 29652.6 \wavn , we observe depletion centered at
19629.4 \wavn  (Fig \ref{fig:depl_fano}a,b).  
The depth is about half of the steady ion signal which demonstrates 
that the Fano type feature is quantitatively depleted. 
The measurements presented in
Figs. \ref{fig:depl_3dn0n1} and
\ref{fig:depl_fano} demonstrate that the lines of series 1 and 2, 
the transitions to the vibrationally ground and symmetric stretch
excited \thrdE\ states  at  17333.3 and 19629.4
\wavn , respectively, and the Fano-type peak at 29652.6
\wavn\  originate from  the vibrationless \twop (0,0,0,0)
metastable state of \DDD . 
These transitions are shown in the level scheme Fig. \ref{fig:levelscheme}  
with the common lower state labelled by I.  

\subsection{Depletion measurements of vibrationally excited \twop\ states}

\begin{figure} [h]
\psfig{figure=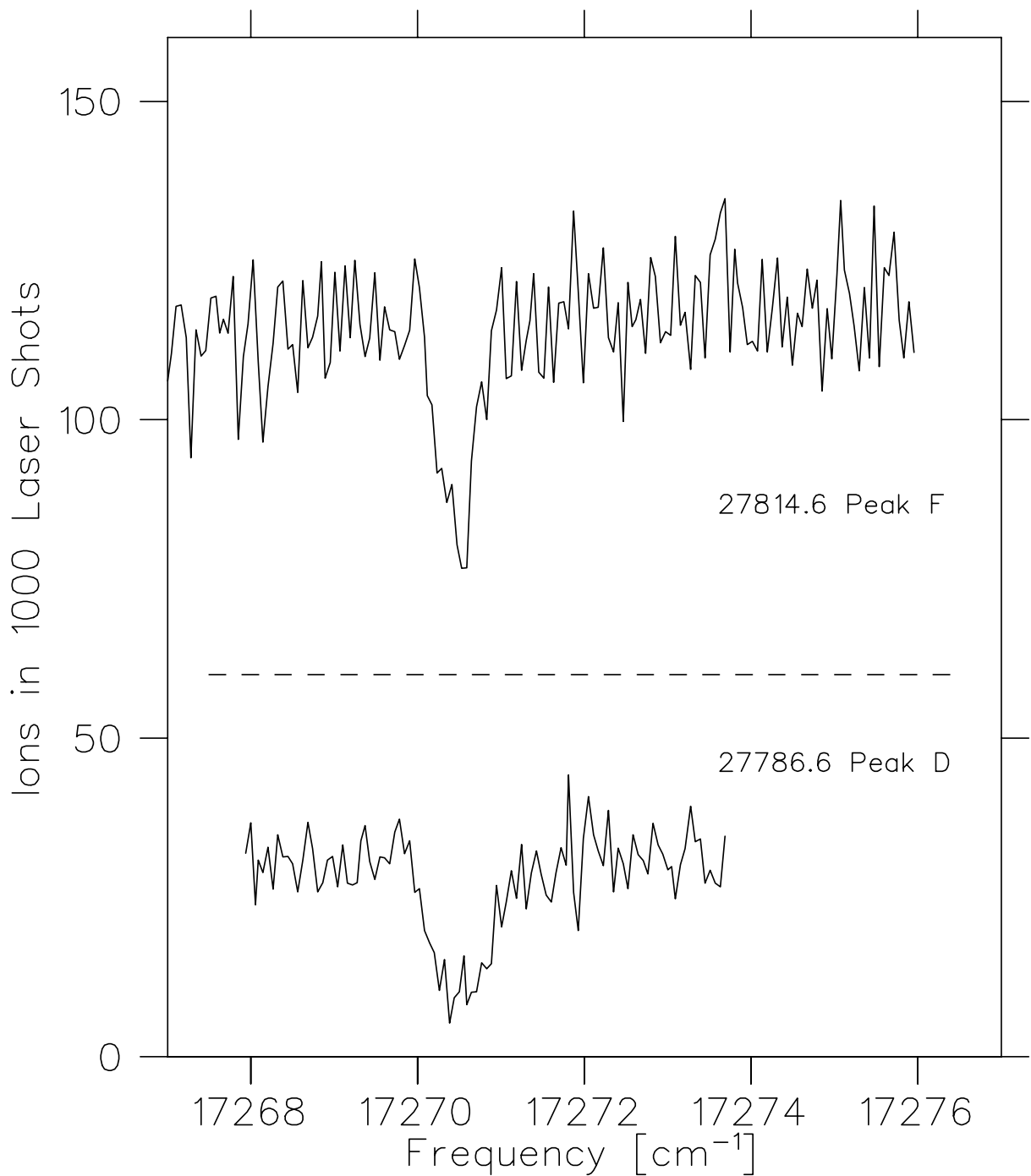,width=16cm}
\caption [autoionizing lines n=7,8]
 {\label{fig:n7n8} 
   Autoionizing states excited by vibrationally diagonal single-photon 
   excitation of vibrationally excited \DDD\ \twop\ molecules 
   in the (a) 27230 \wavn\ to 27350 \wavn\   and the 
   (b) 27760 \wavn\ to 27840 \wavn\ frequency range. 
    The positions of the n=7 and n=8 lines of the Rydberg series 3 to 5 
   (see Table \ref{tb:Series}) are indicated by  open circles. 
 }
\end{figure}

The objective is to determine the core  excitation of the 
autoionizing  d-- and s--Rydberg states of series 3 and 4 converging to
a limit of  29535.1 \wavn , and  the d--states of series 5 converging to
29547.8 \wavn .
In Fig. \ref{fig:n7n8}, vibrationally diagonal transitions 
to autoionizing states in the 27230 \wavn\ to 27350 \wavn\   and the 
 27760 \wavn\ to 27840 \wavn\ energy range are shown.  
The positions of the n=7 and n=8 lines of series
3 to 5 calculated from the series limits and quantum defects in Table
\ref{tb:Series} are  indicated by the open circles  in  Figure \ref{fig:n7n8}. 
As discussed previously \cite{MMRB97}, the thresholds for vibrational 
autoionization of Rydberg states calculated from the molecular
constants of   \DDDp\  \cite{ACCK94} are n=8 and n=7 for a core excitation of 
one quantum in the degenerate and symmetric stretch mode, respectively.
This should allow us to determine the
core excitation of the different series from the
lowest observed principal quantum number lines.
The peaks F, D and G are the n=8 members  of series 3,4, and 5, respectively. 
In the n=7 range, the assignment is not unique. Peak B corresponds 
 to n=7 of series 5. Peak A could belong to series 4. The peak C cannot be 
assigned to  any of the series in Table \ref{tb:Series}.  

\begin{figure} [h]
\psfig{figure=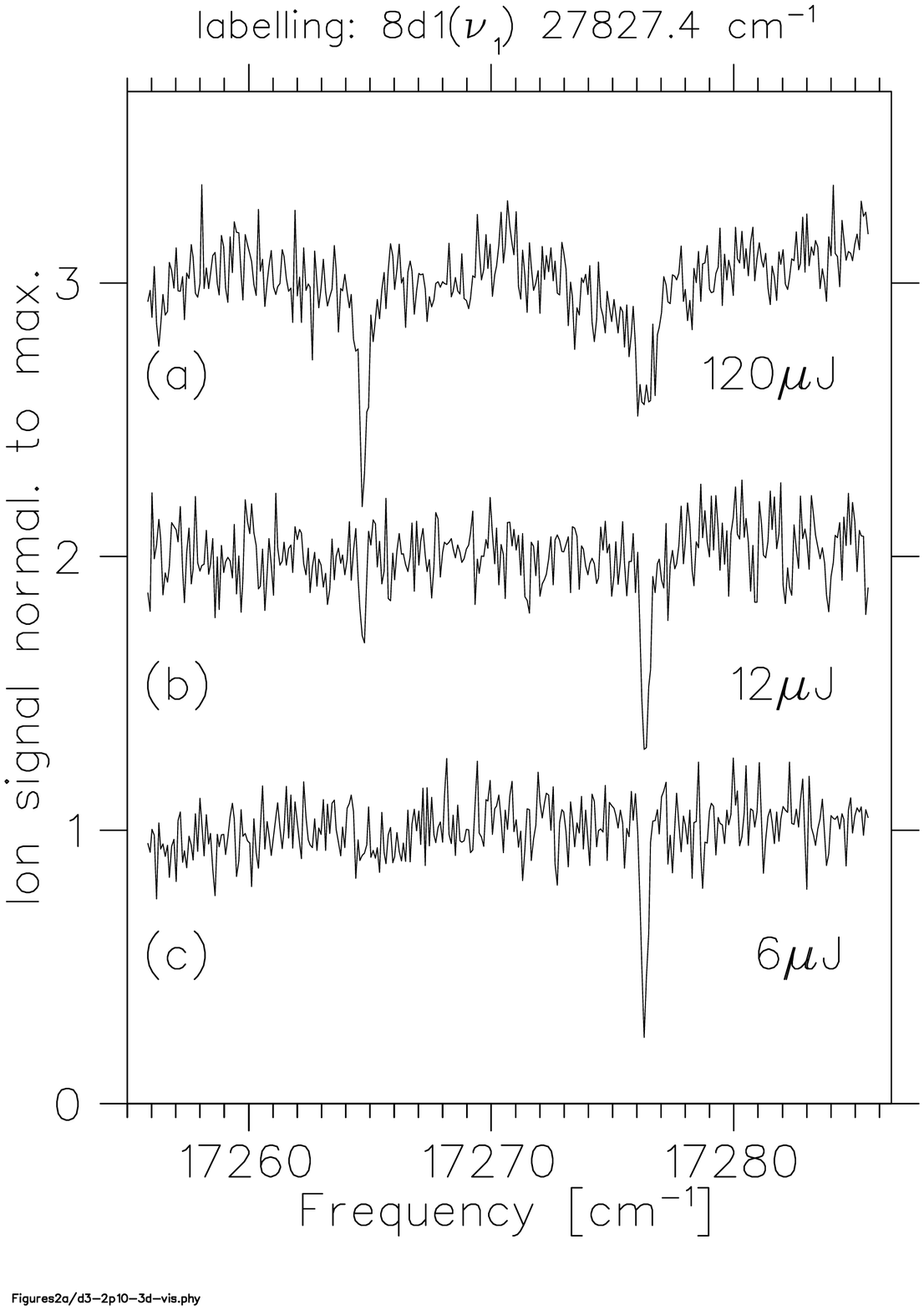,width=16cm}
\caption [depletion n=7,8 nu1]
 {\label{fig:depl_n7n8_nu1} 
   Depletion spectra with the labelling laser tuned to the 
   peaks A,B,C,E, and G of fig. \ref{fig:n7n8}. The excitation laser was 
   scanned in the 17263.5 to 17265.5 \wavn\ spectral range. 
   For clarity of presentation, the spectra are shifted vertically 
   with respect to each other, and separated by a dashed line 
   which indicates the zero-point. 
 }
\end{figure}

\begin{figure} [h]
\psfig{figure=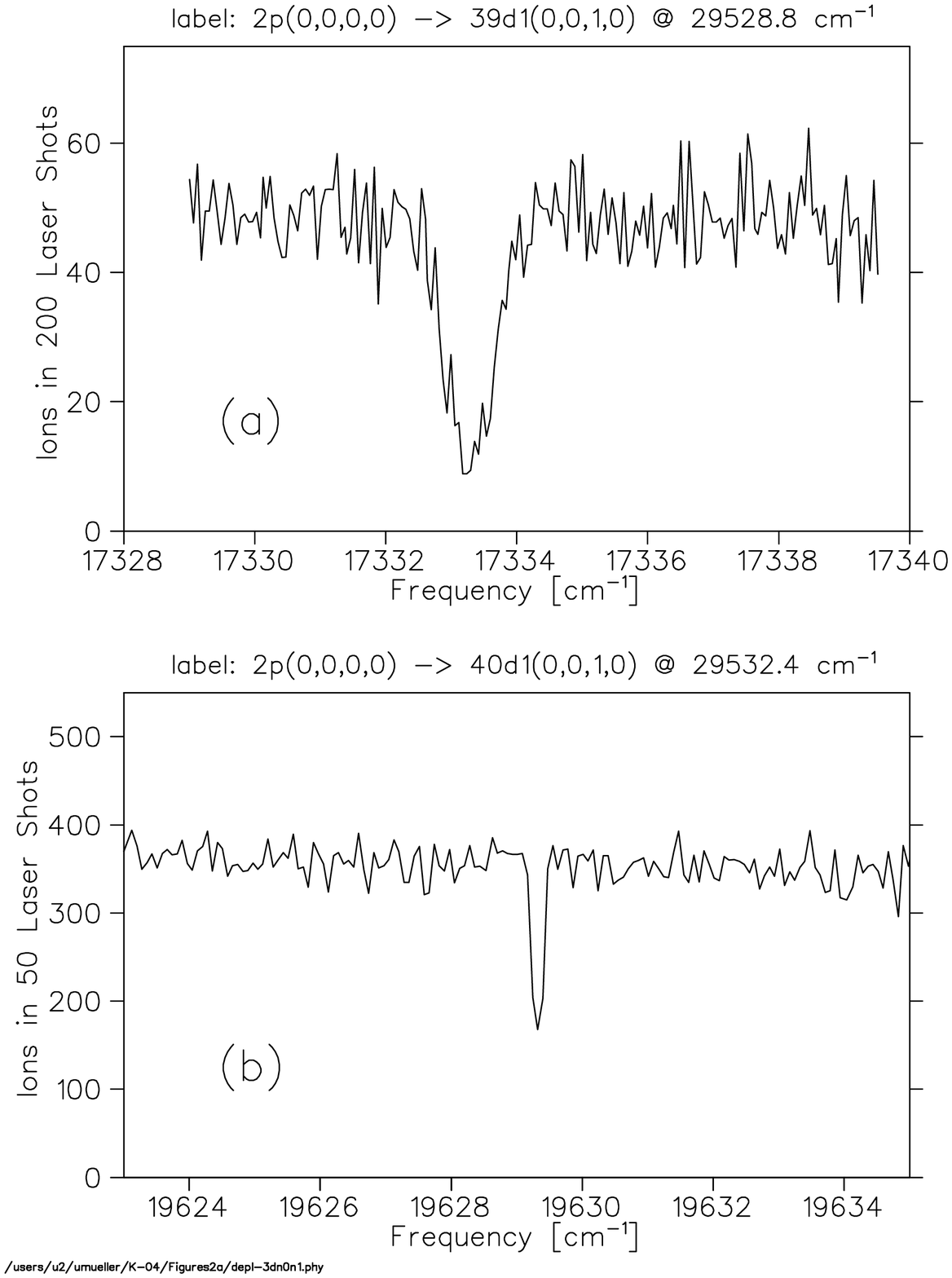,width=16cm}
\caption [depletion n=7,8 nu2]
 {\label{fig:depl_n7n8_nu2} 
   Depletion spectra with the labelling laser tuned to the 
   peaks D and F of fig. \ref{fig:n7n8}. The excitation laser was 
   scanned in the 17268 to 17274 \wavn\ spectral range. 
   For clarity of presentation, the spectra are shifted vertically 
   with respect to each other, and separated by a dashed line 
   which indicates the zero-point. 
 }
\end{figure}

Therefore, we performed
depletion experiments with the labelling laser tuned to the peaks
A to G,  and the excitation laser operating with Rhodamine 6G.
The results are presented in Figs. \ref{fig:depl_n7n8_nu1} and
\ref{fig:depl_n7n8_nu2}. Labelling the peaks A,B,C, E, and G we
 found depletion dips at  17264.5 \wavn . On the other hand, labelling the 
peaks D and F, we clearly observed  dips 
at 17270.5 \wavn . We cross-checked and found  
that the lines   D and F and the lines A,B,C, E, and G are excited from the
common lower states  II and III, respectively, as shown in the  level scheme
Fig. \ref{fig:levelscheme} . None of the features in the left part of Fig.
\ref{fig:n7n8} arises from state II. The peaks 
D and F are the lowest principal quantum number lines  of series 3 and 4 
and the autoionization threshold of these series is n=8.   
It follows that the ion cores of the Rydberg states of 
series 3 and 4  and of the common lower state (II in Fig. \ref{fig:levelscheme}) 
are in the first  degenerate mode excited vibrational state (\nuone =1,
\nutwo =2).  The assignment of the  peaks A, C, and E in  Fig. \ref{fig:n7n8} 
will be discussed below. 

\begin{figure} [h]
\psfig{figure=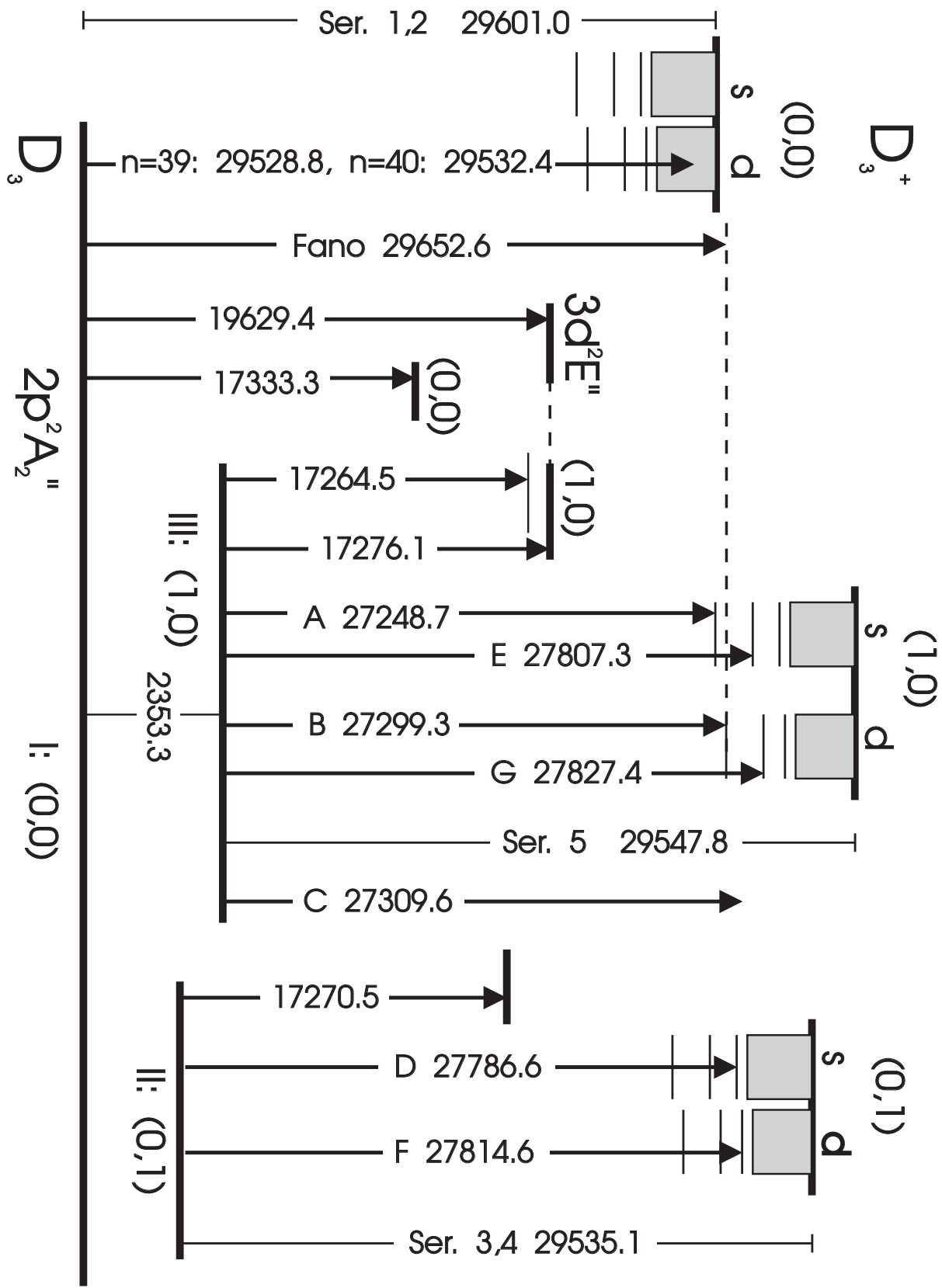,width=16cm}
\caption [depletion n=8, nu1, power dependence]
 {\label{fig:d3-2p-3d-vis} 
   Depletion spectra with the labelling laser tuned to  
   peak G  of fig. \ref{fig:n7n8}. The excitation laser was 
   scanned in the 17255 to 17285 \wavn\ spectral range. 
   In spectrum a, the pulse energy of the excitation laser was 
   120 $\mu$J. In the spectra b and c, the excitation laser was attenuated 
   by factors of 10 and 20 respectively using neutral density filters. 
   The ion signal was normalized. 
   For clarity of presentation, the curves are shifted vertically 
   by one unit with respect to each other. 
 }
\end{figure}

\begin{figure} [h]
\psfig{figure=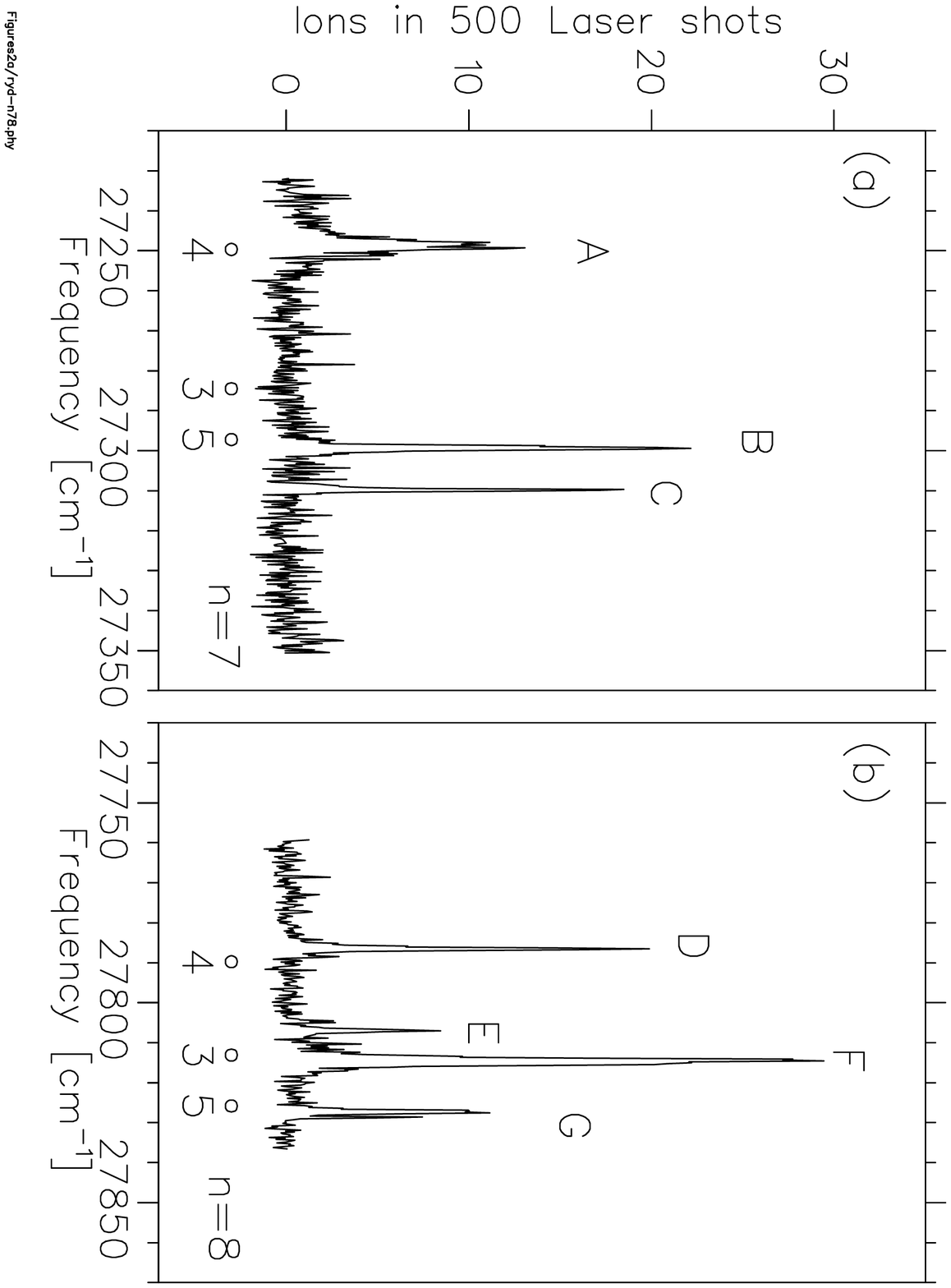,width=16cm}
\caption [level scheme]
 {\label{fig:levelscheme} 
   Level scheme containing states and transitions investigated 
   by depletion spectroscopy in this paper. 
   The vibrational levels are labelled by the 
    quantum  numbers (\nuone ,\nutwo ). The ionization limits 
   of series 1 to 5 (see Table \ref{tb:Series}) 
   are indicated by the thin vertical lines. 
  The pumped transitions are shown by thick vertical lines 
  with arrowheads. 
  Transitions in the ultraviolet range were used for labelling, those 
  in the visible spectral range for depletion. 
 }
\end{figure}

In the measurement presented in Fig. \ref{fig:d3-2p-3d-vis},
 peak G (n=8 of series 5) was excited by the labelling laser 
and the  excitation
laser was tuned in the  17256 \wavn\ to 17285 \wavn\ range. In 
spectrum (a) of Fig.  \ref{fig:d3-2p-3d-vis}, the pulse energy of the
excitation laser was 120 $\mu$J. Spectra (b) and (c) of Fig. 
\ref{fig:d3-2p-3d-vis} were recorded with the excitation laser attenuated by
factors of 10 and 20 respectively using neutral density filters. 
We observe two depletion features at 17264.5 \wavn\ and at 17276.1 \wavn\ 
respectively.   At high pulse energy of the excitation laser, the feature at
17264.5 \wavn\ is comparatively narrow and in the center of the dip, the ion
signal is almost completely  depleted.  
This feature  disappears at low pulse energy.  The dip at  17276.1 \wavn\ 
appears broad and shallow with a width of about 1 \wavn\ FWHM at high 
 pulse energy.  At a laser energy as
low as  6 $\mu$J, the depletion of the ion signal is almost complete and the
width  is reduced by a factor of 3 to about 0.3 \wavn . 
This shows that the 
transition moment of the line at 17276.1 \wavn\ is by at least 
one order of magnitude stronger than that of the line at 17264.5 \wavn .

\subsection{Symmetric Stretch Frequency of the \DDD\ \twop\ State}

In the level scheme Fig. \ref{fig:levelscheme}, the  
transitions which were found by the depletion technique 
to originate from the lower states labelled
 by I, II, and III are indicated by  vertical lines. 
 As already discussed, state I was identified to be the vibrationless 
 metastable state \twop (0,0,0,0) . 
State II was found to be the degenerate mode (\Ep )
excited \twop\ state. 
We notice that the difference between the Fano-type feature at 29652.6 \wavn\
and the \twop (0,0,0,0)$\rightarrow$\thrdE (1,0,1,0) transition
  at 19629.4 \wavn\ is 10023.2 \wavn . 
This value is in (almost too)
perfect  agreement with the difference between the transitions at 
27299.3 \wavn\ (peak
B in Fig. \ref{fig:n7n8})  and  at 17276.1 
\wavn\   (Fig. \ref{fig:d3-2p-3d-vis})  
both originating from state III . We do not find any other combination
differences between the observed transitions which would coincide within the
experimental accuracy. This leads us to conclude that the 
 upper states connected by  the dashed horizontal lines in
Fig. \ref{fig:levelscheme} are identical.  As a consequence, 
the separation between the  states I and III calculated 
via the \thrdE (1,0,1,0)
intermediate state  is found to be 2353.3 \wavn . This value is extremely close
to the first  symmetric stretch vibrational frequency of the \DDDp\ ion. We
conclude that  state III is the symmetric stretch excited \twop (1,0,0,0) state. 
It follows that the previous assignment \cite{MMRB97} 
of series 5 to converge to
a symmetric stretch  excited \DDDp\ state is correct. 

In order to assign the peaks A, C, and E  
we calculated the principal quantum numbers and quantum defects of all the 
features  in Fig. \ref{fig:n7n8} which were found to have the symmetric stretch 
excited \twop (1,0,0,0,) state as the lower state. 
We used the 29547.8 \wavn\ limit of series 5 which converges to a symmetric
stretch excited \DDDp\ state. The values are listed  in Table \ref{tb:Peaks}.
The peaks B and G with n=7, $\delta$
=0.0143  and n=8, $\delta$ =0.0138 respectively have already been 
recognized to
be d states  of series 5 and are listed for completeness in Table \ref{tb:Peaks}.
Based on the quantum defect of $\delta$=0.060,  peak E is assigned to the 
8s state with symmetric stretch excited core. 

The quantum defect  $\delta$=0.092 of peak A is significantly larger than the 
typical values for s--states  (c.f. tab. \ref{tb:Series}) and 
we have to understand the peak width of 3 \wavn\ FWHM which is appreciably
larger than that of the other features in Fig. \ref{fig:n7n8}.  
The  peak
center at 27248.7 \wavn\ is extremely close to the  energy difference of
27247.7 \wavn\ between the symmetric stretch excited 
\twop (1,0,0,0) state and the first ionization limit of the
vibrationless \twop (0,0,0,0)  ground state. Transitions between the 
\twop (1,0,0,0) state and the vibrationless  s-- and d-- states of 
series 1 and 2 are optically allowed, but the Franck-Condon factors are  
very small. Feature A could  arise from an interference between the 
vibrationally diagonal transition from the \twop (1,0,0,0) state to the 
symmetric stretch excited 7s state and non-diagonal transitions to
vibrationless high principal quantum number   s-- and d--  states of  series
1 and 2.

Based on  the quantum defect of $\delta$=-0.002, 
peak C was explained in a
previous study \cite{MMRB97} by an interloper of a g--series with symmetric
stretch excited core.  Another explanation is a vibrationally non-diagonal
transition to a  5s (2,0,1,0) state with two quanta of excitation in the
symmetric stretch mode. The series limit listed in Table \ref{tb:Peaks} was
calculated from the molecular constants of
\DDDp\ determined by Amano et al. \cite{ACCK94}. The  quantum
defect of  $\delta$=0.0566 is in line with the
values generally observed for low principal quantum number s--states.

\subsection{Comparison between Experimental and Theoretical Results}

\begin{figure} [h]
\psfig{figure=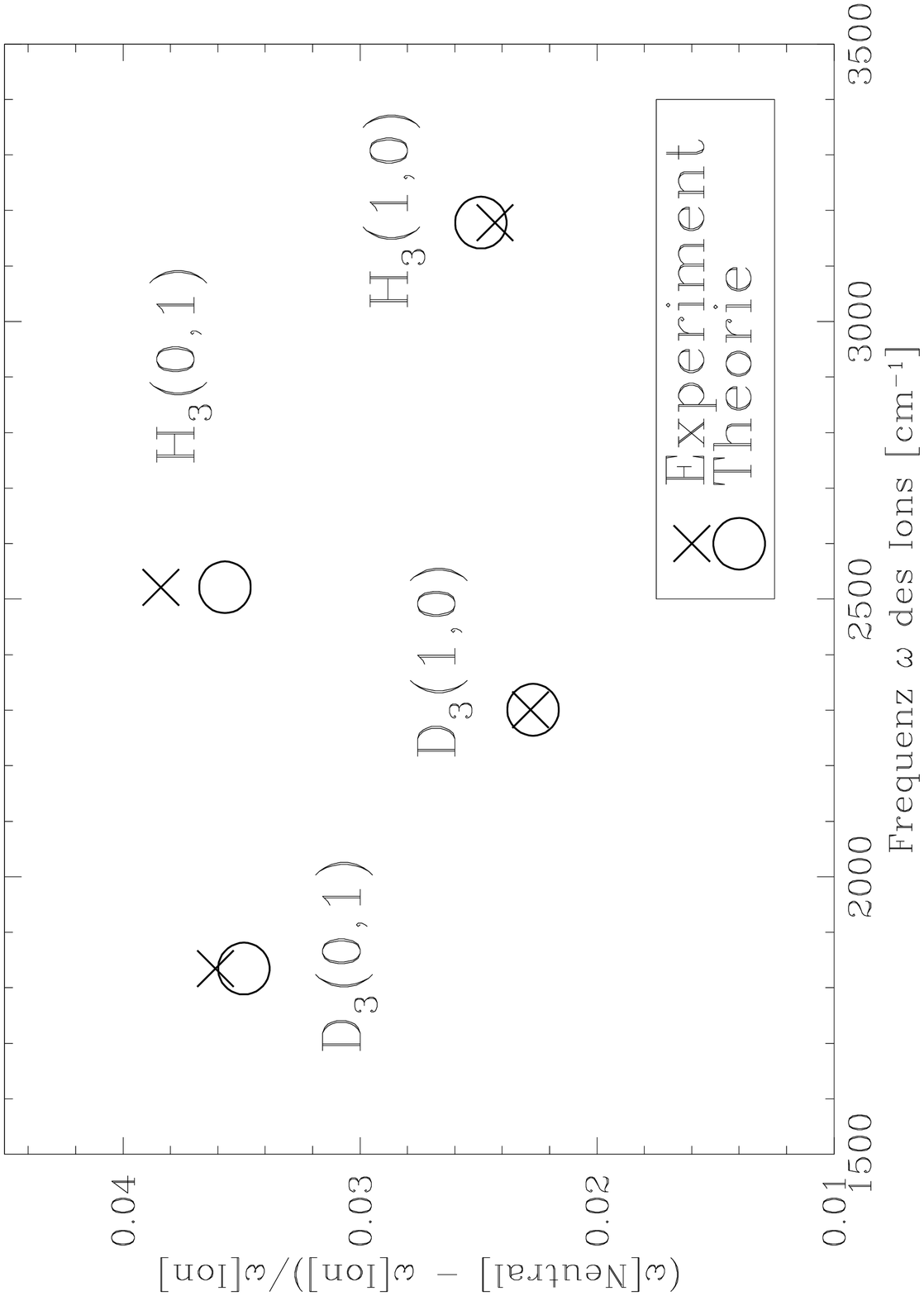,width=16cm}
\caption [viblevels]
 {\label{fig:viblevels} 
   Differences between the vibrational frequencies  of the (neutral) 
   \twop\ states  
   of \DDD\ and \HHH\ and those of the \DDDp\ and \HHHp\ ions. 
   The differences  were normalized to the vibrational frequencies 
   of the corresponding ions. The vibrational states are labelled by the 
    quantum  numbers (\nuone ,\nutwo). \\
 }
\end{figure}

In Table \ref{tb:VibSep}, the 
 energies of the  lowest vibrational states  of  \HHH\ \twop\ and
\DDD\ \twop\ determined by theoretical and experimental investigations 
are listed. 
The values are given relative to the vibrationless
states. 
The theoretical data are  calculated from the  {\it ab initio}
potential energy surface by Peng et al.\cite{PKKW95}  using the procedure
described in Section 3.  For comparison, the vibrational frequencies of the 
\HHHp\ and \DDDp\ ions determined by Majewski et al. \cite{MMSW94} 
and by  Amano et al. \cite{ACCK94} are included in Table \ref{tb:VibSep}. 
The differences between the  vibrational frequencies of the
neutrals and the ions are
quite large which shows the  valence character of the  \twop\ state. The
use of  {\it ab initio} potential energy surfaces is mandatory in order to
calculate the vibrational levels to a satisfactory accuracy.  A treatment based
on a quantum defect, which is independent of the nuclear coordinates, is
condemned to fail. 
 The experimental results 
for \DDD\ were measured in this  and in a previous investigation from 
this laboratory \cite{MMRB97}, and the data for   \HHH\ 
were taken from  Ketterle et al. \cite{KMW89}. The value of  3255.4 \wavn\
for the symmetric stretch frequency of the \HHH\ \twop\ state measured by Lembo
and Helm \cite{LH89} is in close agreement with the value of 3255.38 \wavn\  
measured by Ketterle et al. \cite{KMW89}. 
The uncertainties of the experimental data which are estimated to
be $\pm$0.3 \wavn\ for \DDD\ and $\pm$0.1 \wavn\ for \HHH ,  
are generally much smaller 
than the  differences between experiment and theory 
which are of the order of a few \wavn , the largest being 6.6
\wavn\   in case of the  \Ep -- vibration of  \HHH\ \twop . As discussed above, 
the numerical procedure for the calculation 
of the eigenvalues is converged to an accuracy better than 0.25 \wavn . 
Therefore, the differences between experiment and theory must be 
attributed to the  potential energy surface which  was fitted to
the {\it ab initio} points with an average deviation of 5 \wavn\ and  
 a root mean square deviation of 170 \wavn\ \cite{PKKW95}. 
The shape of the fitted PES in the region occupied by the low vibrational states
is apparently quite good. 
The  observed 
differences between the vibrational frequencies of the neutrals and the ions
are in very good agreement with  the theoretical data as
visualized in  Fig. \ref{fig:viblevels} where we present the 
differences between the vibrational frequencies of the 
\twop\ states and those of the  corresponding ions,  normalized to
the  frequencies of the ions for easier
comparison. The measured symmetric stretch 
frequencies of the 
\twop\  states of \DDD\ and \HHH\ are  roughly 2 \% higher than those of 
the \DDDp\ and  \HHHp\ ions, respectively, which is very well reproduced
by the calculated data. 
In the case of the degenerate mode vibration,  
the relative differences between  neutrals and 
ions  are almost twice as large. 
This additional resistance against asymmetric deformation induced by  
the 2p$_z$--electron is  explained by the theoretical data, although the
total effect is slightly underestimated.

\section{Conclusions}

We investigated the vibrational frequencies of the
\twop\ metastable state  and the \thrdE\ state of the triatomic deuterium
molecule \DDD\ by a depletion type  double resonance technique. 
We found a  separation of 2296.1 \wavn\ between the vibrationless 
\DDD\ \thrdE (0,0,1,0) and the symmetric stretch excited \thrdE (1,0,1,0) 
state.
The 2353.3 \wavn\ symmetric stretch frequency of the \twop\ state determined in 
a previous investigation from this laboratory was confirmed. 
We compared the experimental results of the \twop\ state of \DDD\ and
\HHH\  with theoretical data calculated from an {\it ab initio} potential energy
surface. 
The theoretical  data describe very well 
the differences  between the vibrational frequencies of the ions and the 
neutrals  which demonstrates the  quality  of the 
{\it ab initio} potential energy surface. 

\section{Acknowledgements}
We are greatly indebted to Prof. H. Helm and Prof. Ch. Schlier for 
their support, for continuous 
encouragement during the course of this work, and for critical reading 
of the manuscript. 
Thanks must go to S. Kristyan for 
 providing us with the fitted potential energy surface. 
It is a pleasure to acknowledge  technical assistance by 
U. Person  during the 
construction and operation of the apparatus.  
This research was financially supported by 
Deutsche Forschungsgemeinschaft through its SFB 276.

\newpage
\section{Tables}
\vspace*{1.5cm}

\begin{table}[ht]
\caption{\label{tb:Series} 
Rydberg series observed following laser-photoexcitation
of metastable \DDD\ \twop\ molecules. }
\par
\vspace*{0.5cm}
\begin{tabular}{||c|r|r|c|c|c||}
\hline
Nr. & \multicolumn{1}{c|}{$\rm E_{lim}$[\wavn ] $^a$} 
    & \multicolumn{1}{c|}{$\delta$} & observed n  
    & \twop\ $^b$  & upper state $^b$\\ \hline
1 & 29601.0  & 0.015(8)  & 20-80 & (0,0,0,0) & nd1 (0,0,1,0)\\ \hline
2 & 29601.0  & 0.08(1)   & 22-31 & (0,0,0,0) & ns1 (0,0,1,0)\\ \hline
3 & 29535.1  & 0.014(1)  & 8-71  & (0,1,0,1) & nd1 (0,1,1,1)\\ \hline
4 & 29535.1  & 0.07(6)   & 8-31  & (0,1,0,1) & ns1 (0,1,1,1)\\ \hline
5 & 29547.8  & 0.015(1)  & 7-46  & (1,0,0,0) & nd1 (1,0,1,0)\\ \hline
\end{tabular} 
\par
 $^a$ The statistical uncertainty of the fit procedure is 
smaller than the 0.2 \wavn\ systematic uncertainty 
of the wavelength calibration 
\par
$^b$ The notation for the rovibrational state is (\nuone,\nutwo,N,K)
\end{table}

\vspace*{1.5cm}

\begin{table}[ht]
\caption{ \label{tb:Peaks}  Assignment of the peaks 
  in fig. 7 .}
\par
\vspace*{0.5cm}
\begin{tabular}{||c|r|c|r|c|r||}
\hline
Peak	    &position	& core vibration	& $\rm E_{lim}$	 & n	& $\delta$ \\ \hline
A	       & 27248.3	& (1,0)	& 29547.8$^a$	& 7	& 0.092  \\ \hline
B	       & 27299.3	& (1,0)	& 29547.8$^a$	& 7	& 0.0143  \\ \hline
C	       & 27309.8	& (1,0)	& 29547.8$^a$	& 7	& -0.0021 \\ \hline
E	       & 27807.3 & (1,0)	& 29547.8$^a$	& 8	& 0.0600  \\ \hline  
G	       & 27827.4 & (1,0)	& 29547.8$^a$	& 8	& 0.0138   \\ \hline \hline
C	       & 27309.8	& (2,0)	& 31800.0$^b$	& 5	& 0.0566  \\  \hline
\end{tabular} 
\par
 $^a$ Series limit of series 5 in table 1 \\
 $^b$ (\nuone = 2, \nutwo =0) ionization limit with respect to  the 
     \twop (\nuone = 1, \nutwo =0) state of \DDD\ calculated from the ionization 
limit of series 5 and the molecular constants of \DDDp\ \cite{ACCK94}.
\end{table}

\vspace*{1.5cm}

\begin{table}[ht]
\caption{ \label{tb:VibSep}  Vibrational Energies of the 
  \twop\ states of \HHH\ and \DDD\ compared to those of the corresponding ions. }
\par
\vspace*{0.5cm}
\small
\begin{tabular}{||c|c|c|c|c|c||c|c|c|c|c||}
\hline
vib.	& \HHHp\ & \multicolumn{4}{|c||}{\HHH\twop\ }& 
  \DDDp\ $^e$  &  \multicolumn{4}{|c||}{\DDD\twop\ } \\ \hline
state$^a$	& ion$^b$ &th. $^c$ & th.-ion & exp. $^d$& exp.-th. &
  ion $^e$  & th.	$^c$ & th.-ion & exp. $^f$  & exp.-th. \\ \hline
(0,1)$^1$&2521.416&2611.7& 90.3&2618.34& 6.6&1834.674&1898.8& 64.1&1900.9&2.1 \\ \hline
(1,0)$^0$&3178.177&3257.6& 79.4&3255.38&-2.2&2300.843&2353.3& 52.5&2353.3&0.0 \\ \hline
(1,1)$^1$&4778.228&4951.9&173.7&       &    &3530.385&3650.1&119.7&      &  \\ \hline
(0,2)$^0$&4997.920&5181.5&183.6&       &    &3650.658&3777.1&127.1&      &  \\ \hline
(0,2)$^2$&5554.029&5734.1&180.0&      	&    &4059.470&4182.2&122.7&      &  \\ \hline
(2,0)$^0$&6262.213&6426.6&164.4&     	 &    &4553.792&4661.6&107.8&      &  \\ \hline
\end{tabular} 
\par $^a$ the vibrational states (\nuone ,\nutwo)$^l$ are labelled by  
the number of quanta in the symmetric stretch mode \nuone , the degenerate mode 
\nutwo , and the vibronic angular momentum l. 
The energy of the  vibrationless  (0,0)$^0$ state was  set to zero. 
\par $^b$ \ molecular constants from ref. \cite{MMSW94} .
\par $^c$ this work, data calculated from the fitted potential energy
  surface of ref. \cite{PKKW95}.
\par $^d$ \ experimental data, ref. \cite{KMW89}.
\par $^e$  \ molecular constants from ref. \cite{ACCK94} .
\par $^f$ \ experimental data, this work and ref. \cite{MMRB97}.
\end{table}

\clearpage
\section {Figure captions} 
\par
\par
\par
\par
\par
\par
\par
\par

\begin{thebibliography}{99}

\bibitem{NJ82}
  Ch. Nager and M. Jungen, Chem. Phys.{\bf 70}, 189 (1982) 
\bibitem{PKKW95} Z. Peng, S. Kristyan, A. Kuppermann, and J. S. Wright, 
  Phys. Rev. A. {\bf 52}, 1005 (1995)
\bibitem{HerzbI} I. Dabrowski and G. Herzberg, 
    Can. J. Phys.  {\bf 58}, 1238 (1980)
\bibitem{HerzbII} G. Herzberg and J. K. G. Watson, 
    Can. J. Phys.  {\bf 58}, 1250 (1980)
\bibitem{HerzbIII} G. Herzberg, H. Lew, J. J. Sloan,  and J. K. G. Watson, 
    Can. J. Phys.  {\bf 59}, 428 (1981)
\bibitem{HerzbIV} G. Herzberg, J. T. Hougen, and J. K. G. Watson, 
    Can. J. Phys.  {\bf 60}, 1261 (1982)
\bibitem{HLCH89}
  H. Helm, L. J. Lembo, P. C. Cosby, and D. L. Huestis, 
   Fundamentals of Laser Interactions II, 
   Lecture Notes in Physics (Ed. F. Ehlotzky),
   Springer (1989), p. 264
\bibitem{Dev69}
  F. M. Devienne, C. R. Acad. Sci. Paris B  
  {\bf 267}, 1279 (1968) and {\bf 268}, 1303 (1969)
\bibitem{GP83}
  G. I. Gellene and R. F. Porter,  
   J. Chem. Phys. {\bf 79}, 5975 (1983)
\bibitem{GK84}
  J. F. Garvey and A. Kuppermann, 
   Chem. Phys. Lett. {\bf 107}, 491 (1984)
\bibitem{He86} H. Helm, Phys. Rev. Lett. {\bf 56}, 42 (1986)
\bibitem{He88} H. Helm, Phys. Rev. A {\bf 38}, 3425 (1988)
\bibitem{DKMW88}
  A. Dodhy, W. Ketterle, H.-P. Messmer, and H. Walther, 
    Chem. Phys. Lett. {\bf 151}, 133 (1988)
\bibitem{KMW89}
  W. Ketterle, H.-P. Messmer, and H. Walther, 
    Europhys. Lett. {\bf 8}, 333 (1989)
\bibitem{LPH88}
  L. J. Lembo, A. Petit, and H. Helm, 
    Phys. Rev. A{\bf 39}, 3721 (1989)
\bibitem{LH89}
  L. J. Lembo and H. Helm, 
    Chem. Phys. Lett. {\bf 163}, 425 (1989)
\bibitem{LHH89}
  L. J. Lembo, D. Huestis, and H. Helm, 
    J. Chem. Phys. {\bf 90}, 5299 (1989)
\bibitem{MMRB97}  U. M\"uller, U. Majer, R. Reichle, and M. Braun, 
     JCP {\bf 106}, 7958 (1997)
\bibitem{ACCK94} 
  T. Amano, M.-C. Chan, S. Civis., A. R. W. McKellar, 
  W. A. Majewski, D. Sadovskii, and J. K. G. Watson, 
  Can. J. Phys.  {\bf 72}, 1007 (1994)

\bibitem{mandelshtam:97a}
V.~A. Mandelshtam and H.~S. Taylor, J. Chem. Phys. {\bf 106},  5085  (1997).
\bibitem{tal-ezer:84}
H. Tal-Ezer and R. Kosloff, J. Chem. Phys. {\bf 81},  3967  (1984).

\bibitem{MIT-tab} MIT Wavelength Tables, G. R. Harrison, Cambridge (1969)
\bibitem{Rank59} D. H. Rank in Advances in Spectroscopy I, 
  ed. H. W. Thompson, New York (1959)

\bibitem{neuhauser:95}
D. Neuhauser and M. Wall, J. Chem. Phys. {\bf 102},  8011  (1995).
\bibitem{sutcliffe:91}
B.~T. Sutcliffe and J. Tennyson, Intern. J. Quantum Chemistry {\bf 39},
183 (1991).
\bibitem{tennyson:85}
J. Tennyson and B.~T. Sutcliffe, Mol. Phys. {\bf 56},  1175  (1985).
\bibitem{colbert:92}
D.~T. Colbert and W.~H. Miller, J. Chem. Phys. {\bf 196},  1982  (1992).
\bibitem{salzgeber:97a}
R.~F. Salzgeber, V.~A. Mandelshtam, Ch. Schlier, and H.~S. Taylor, 1997, 
to be submitted.
\bibitem{Hougen62} J. T. Hougen, J. Chem. Phys. {\bf 37},  1433  (1962).
\bibitem{MMSW94} 
 W. A. Majewski, A. R. W. McKellar,  D. Sadovskii, and J. K. G. Watson ,
 Can. J. Phys.  {\bf 72}, 1016 (1994)
%
%

\end{thebibliography}
\end{document}